\documentclass[journal=jacsat,manuscript=article]{achemso}
\usepackage{textcomp,gensymb,amsmath,xcolor,physics,pdfpages}
\newcommand{\bc}{\begin{center}}
	\newcommand{\ec}{\end{center}}

\newcommand{\beq}{\begin{equation}}
\newcommand{\eeq}{\end{equation}}
\newcommand{\beqs}{\begin{eqn*}}
	\newcommand{\eeqs}{\end{eqn*}}
\newcommand{\bq}{\begin{quote}}
	\newcommand{\eq}{\end{quote}}
\newcommand{\beqa}{\begin{eqnarray}}
\newcommand{\eeqa}{\end{eqnarray}}
\newcommand{\beqas}{\begin{eqnarray*}}
	\newcommand{\eeqas}{\end{eqnarray*}}
\newcommand{\bfg}{\begin{figure}}
	\newcommand{\efg}{\end{figure}}

\newcommand{\txb}{} 

\author{Sarthak Das}
\affiliation[IIScECE]
{Department of Electrical Communication Engineering, Indian Institute of Science, Bangalore 560012, India}
\author{Garima Gupta}
\affiliation[IIScECE]
{Department of Electrical Communication Engineering, Indian Institute of Science, Bangalore 560012, India}
\author{Suman Chatterjee}
\affiliation[IIScECE]
{Department of Electrical Communication Engineering, Indian Institute of Science, Bangalore 560012, India}
\author{Kenji Watanabe}
\affiliation[NIMS1]
{Research Center for Functional Materials, National Institute for Materials Science,
	1-1 Namiki, Tsukuba 305-044, Japan}
\author{Takashi Taniguchi}
\affiliation[NIMS2]
{International Center for Materials Nanoarchitectonics, National Institute for Materials Science, 1-1 Namiki, Tsukuba, 305-044 Japan}
\author{Kausik Majumdar}
\email{kausikm@iisc.ac.in}
\affiliation[IIScECE]
{Department of Electrical Communication Engineering, Indian Institute of Science, Bangalore 560012, India}
\title{Highly nonlinear biexcitonic photocurrent from ultrafast inter-layer charge transfer}

\begin{document}

{\abstract Strong Coulomb interaction in monolayer semiconductors allows them to host optically active large many-body states, such as the five-particle state, charged biexciton. Strong nonlinear light absorption by the charged biexciton under spectral resonance, coupled with its charged nature, makes it intriguing for nonlinear photodetection - an area that is hitherto unexplored. Using the high built-in vertical electric field in an asymmetrically designed few-layer graphene encapsulated 1L-WS$_2$ heterostructure, here we report a large, highly nonlinear photocurrent arising from the strong absorption by two charged biexciton species under zero external bias (self-powered mode). Time-resolved measurement reveals that the generated charged biexcitons transfer to the few-layer graphene in a timescale of sub-5 ps, indicating an ultrafast intrinsic limit of the photoresponse. By using single- and two-color photoluminescence excitation spectroscopy, we show that the two biexcitonic peaks originate from bright-dark and bright-bright exciton-trion combinations. Such innate nonlinearity in the photocurrent due to its biexcitonic origin, coupled with the ultrafast response due to swift inter-layer charge transfer, exemplifies the promise of manipulating many-body effects in monolayers towards viable optoelectronic applications.}
\newpage
A superlinear photodetector, where the output electrical signal varies in a superlinear manner with the input light intensity, is attractive for several functional optoelectronic applications, such as, higher harmonics generator, square-law detector, compensator for sublinearity, optical implementation of neural circuits, and several other analog applications. However, there are very few reports to date \cite{klee2015superlinear,ma2016tuning,massicotte2016photo,dandu2021negative} demonstrating a nonlinear photocurrent, and the reported degree of nonlinearity is often weak and highly dependent on the design/device parameter. A nonlinear photocurrent generation from the intrinsic nonlinear absorption of the active material will thus be highly attractive.
\\
\\
Reduced dielectric screening and enhanced quantum confinement lead to a high binding energy of excitonic complexes in monolayers of transition metal dichalcogenides (TMDs). The TMD monolayers thus serve as an excellent test bed to probe many-body effects in complex excitonic states \cite{ye2018efficient,chen2018coulomb,zinkiewicz2021excitonic,barbone2018charge,steinhoff2018biexciton,hao2017neutral,paur2019electroluminescence}. The signatures of two (exciton), three (charged exciton or trion), four (biexciton) or five-particle (charged biexciton) states have been extensively probed through several optical spectroscopic techniques \cite{kallatt2019interlayer, ye2018efficient,chen2018coulomb,zinkiewicz2021excitonic,barbone2018charge,steinhoff2018biexciton,hao2017neutral,paur2019electroluminescence,jadczak2019room,mostaani2017diffusion,rodin2020collective, chatterjee2021probing}. These many-body states along with their different spin and valley configurations of the constituent particles further lead to a nexus of complex states \cite{zinkiewicz2021excitonic,liu2020multipath,li2019emerging}. Of particular interest is the large five-particle state - charge biexciton \cite{ye2018efficient,chen2018coulomb,zinkiewicz2021excitonic,barbone2018charge,chatterjee2021probing}, which dominates the optical response of monolayer WS$_2$ and WSe$_2$ at relatively higher power density because of their strong nonlinearity due to many-body effect\cite{ye2018efficient,chen2018coulomb,zinkiewicz2021excitonic,barbone2018charge,chatterjee2021probing}. The charged biexciton state also exhibits large gate tunability \cite{ye2018efficient,chen2018coulomb,zinkiewicz2021excitonic,barbone2018charge}, electric field induced quantum confined Stark effect \cite{abraham2021anomalous}, and magnetic field induced splitting \cite{ye2018efficient,zinkiewicz2021excitonic}. As the neutral biexcitons can be formed by both ``bright-bright" and ``bright-dark" excitons \cite{ye2018efficient,chen2018coulomb,barbone2018charge,steinhoff2018biexciton,hao2017neutral}, similar configurations of exciton-trion can lead to two different types of charged biexciton formation \cite{zinkiewicz2021excitonic,chatterjee2021probing}. The strong nonlinear light absorption makes both these charged biexcitonic states an interesting system for nonlinear photocurrent generation - an area that has not been explored to date.
\\
\\
In this work, we reveal the configuration of such charged biexcitonic species in WS$_2$ with the help of a combination of one- and two-color excitation spectroscopy. Using an asymmetric van der Waals heterostructure, we demonstrate a strong nonlinear photocurrent generation in self-powered mode (under zero external bias) through photo-absorption by these two types of charged biexciton, followed by their ultrafast ($< 5$ ps) inter-layer transfer.
\\
\\
\section*{Result and Discussion:}
\begin{figure}[!hbt]
	\centering
	\includegraphics[scale=0.5]{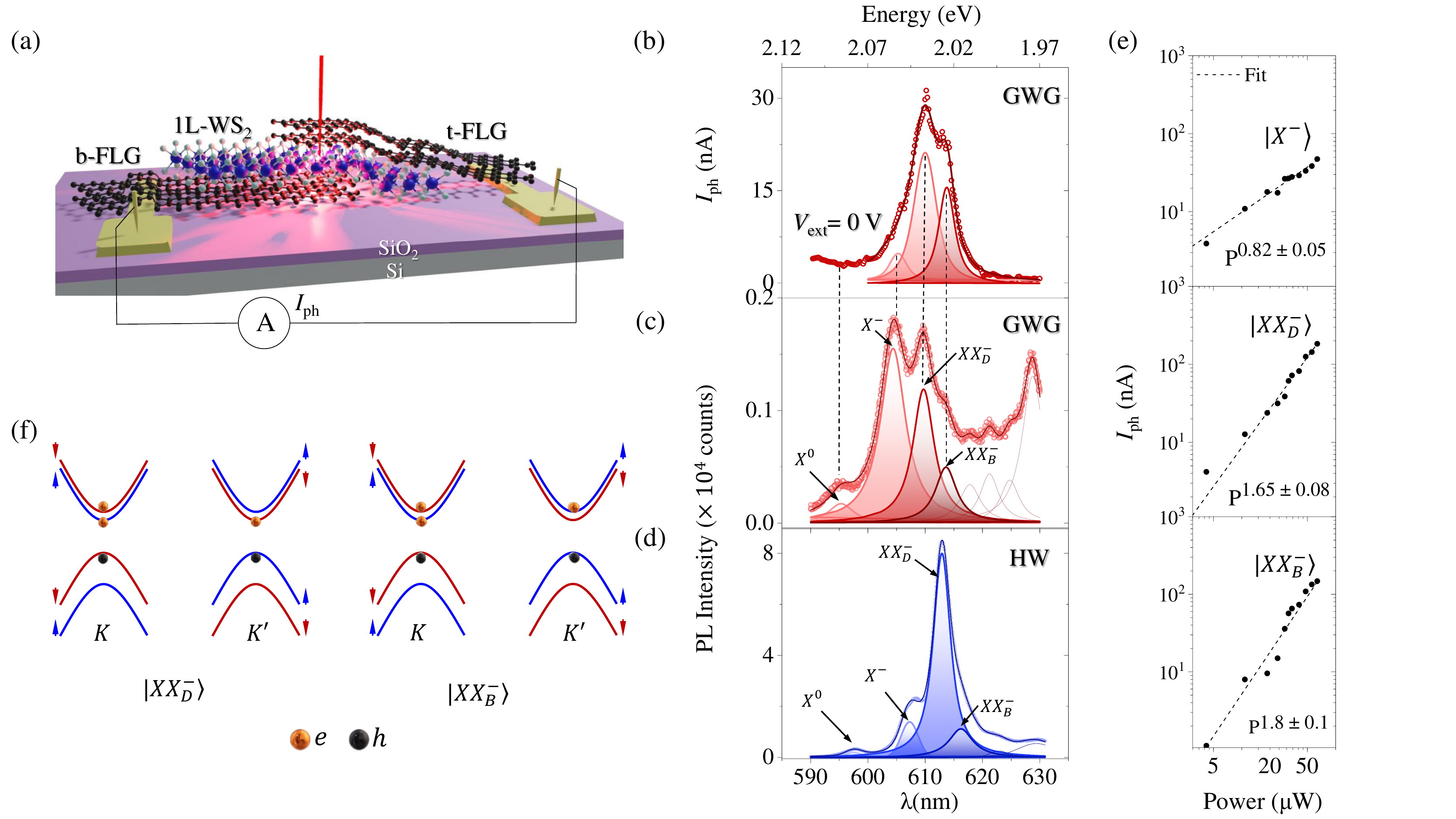}
	\caption{\textbf{Photocurrent from charged biexciton states.}
		(a)  Schematic diagram of the 1L-WS$_2$ device (GWG) encapsulated with few-layer graphene from top (t-FLG) and bottom (b-FLG). The photocurrent ($I_{ph}$) is measured in a short-circuit configuration with zero external bias.
		(b) $I_{ph}$ as a function of the excitation wavelength ($\lambda$) at $T$ = 4.5 K. The spectroscopic response is decomposed into individual trion ($X^-$) and charged biexciton ($XX_{B,D}^-$) peaks.
		(c-d) Photoluminesce (PL) intensity of 1L-WS$_2$ from the GWG portion (c) and WS$_2$ on hBN (HW) portion (d). The vertical dashed lines indicate that the $I_{ph}$ peaks match well with the PL peaks. The small shift between the GWG and HW PL peaks arise from different dielectric environment.
		(e) Optical power dependent $I_{ph}$ for the $X^-$, $XX_{D}^-$ and $XX_{B}^-$ states (symbols). The fitted trend ($I_{ph}\propto P^\alpha$) is shown in dashed lines.
		(f) Schematic representation of the configuration of dark ($XX_{D}^-$) and bright ($XX_{B}^-$) charged biexciton states.
	}\label{fig:Iph}
\end{figure}

\begin{figure}[!hbt]
	\centering
	\includegraphics[scale=0.5]{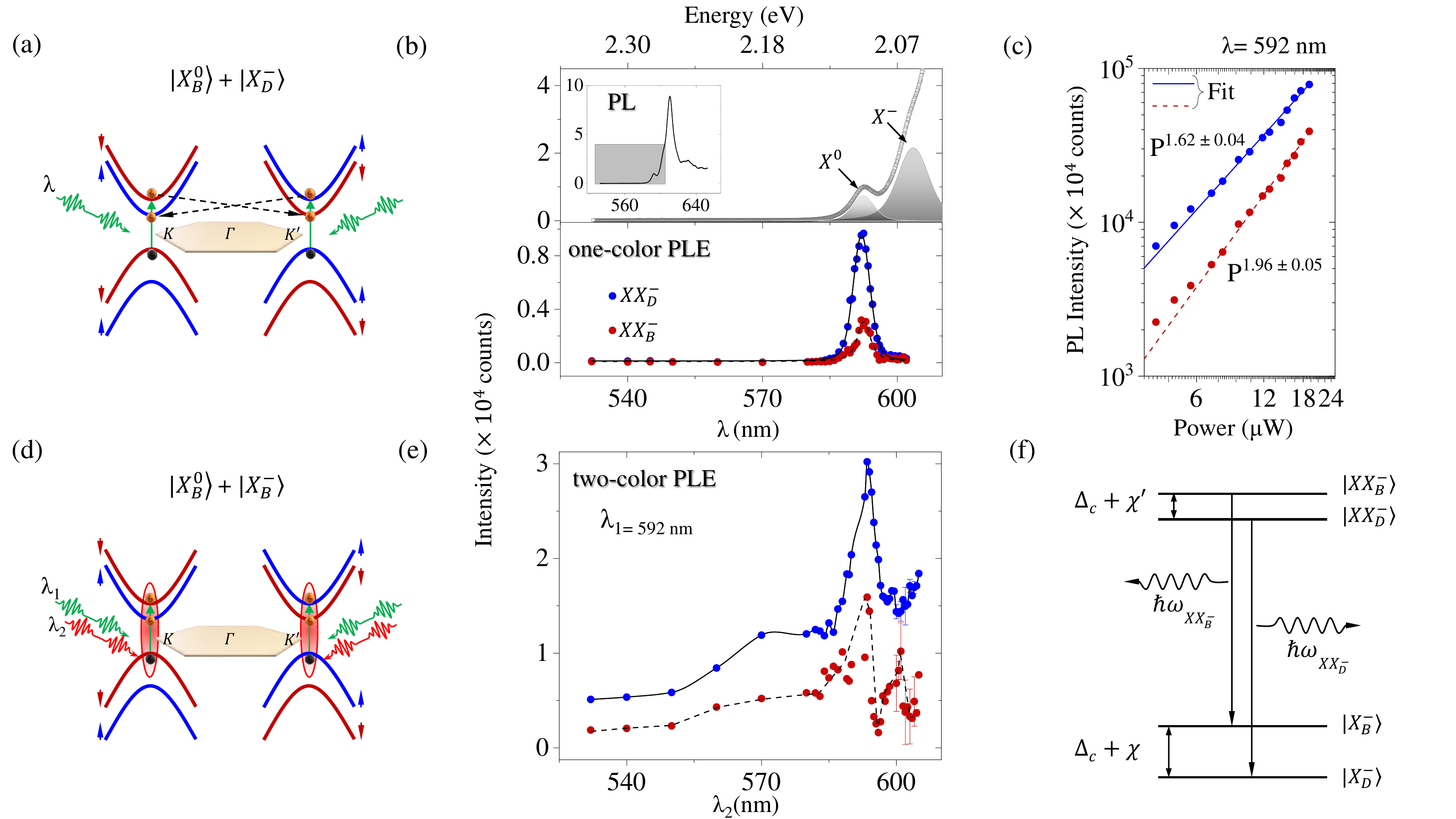}
	\caption{\textbf{One- and two-color photoluminescence excitation spectroscopy of charged biexcitons.}
		(a) Schematic diagram of the one-color photoluminescence excitation (PLE) spectroscopy. The dashed lines indicate the formation of lower energy dark states when the excitation is in resonance with neutral bright exciton.
		(b) Top panel: Zoomed-in PL spectrum for WS$_2$ showing the $X^0$ and $X^-$ peak, with the full spectrum in the inset. Bottom panel: Emission intensity of the $XX_{D}^-$ (blue symbol) and $XX_{B}^-$ (red symbol) states, as a function of the excitation wavelength (energy) in the bottom (top) axis. Both of them show a pronounced emission when excitation wavelength ($\lambda$) matches with the $X^0$. The solid and dashed lines are guide to eye for the  $XX_{D}^-$ and $XX_{B}^-$ one-color PLE intensity, respectively.
		(c) Power dependent PL intensity of the charged biexcitons ($XX_{B,D}^-$) with excitation wavelength at 592 nm ($X^0$ resonance). The solid and dashed lines represent the power law fits.
		(d) Schematic diagram of the two-color PLE, where $\lambda_1$ is kept fixed at $X^0$ resonance, and $\lambda_2$ is scanned. The specific situation of $\lambda_2$ matching the trion state is highlighted.
		(e) PLE intensity for a two-color PLE for $XX_{D}^-$ (blue symbol) and $XX_{B}^-$ (red symbol).
		(f) Transition energy diagram corresponding to the $XX_{B,D}^-$ states. Here $\Delta_c$ and $\chi$ ($\chi^\prime$) denote the spin-orbit splitting in the conduction bands, and the exchange energy of the trion (charged biexciton) state.
	}\label{fig:PLE}
\end{figure}

\begin{figure}[!hbt]
	\centering
	\includegraphics[scale=0.5]{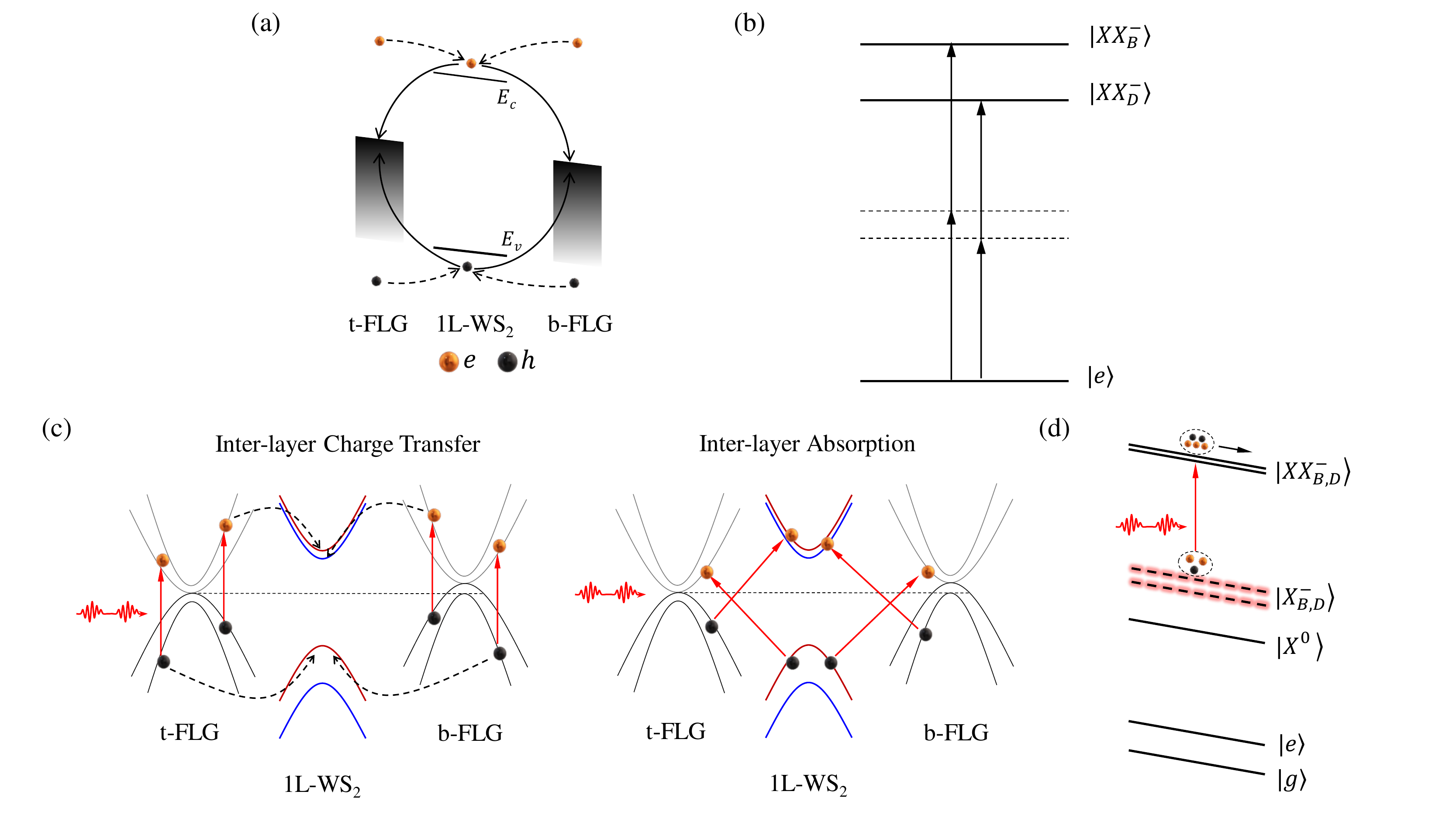}
	\caption{\textbf{Mechanism of $I_{ph}$ generation through charged biexcitons.}
		(a) The light absorbed by top and bottom FLG layers results injection of hot electrons and holes from both the FLG layers to WS$_2$ (dashed arrow). However, due to the large band offset between WS$_2$ and FLG on both sides of WS$_2$, these hot carriers can move towards either direction (solid arrow), hence suppress the $I_{ph}$.
		(b) Schematic of the two-photon absorption (TPA) process where energy conservation indicates that in the TPA, the photo-absorption must occur at an energy equal to $0.5(\hbar\omega_{XX_{B,D}^-} + \hbar\omega_{X_{B,D}^-})$. The dashed lines indicate the virtual states.
		(c) Left panel: Hot electrons and holes generated in the top and bottom FLG layers are injected into 1L-WS$_2$. These non-equilibrium hot electrons and holes occupy the bands of both valleys in WS$_2 $ that eventually prepare the initial bright and dark trion state. Right panel: Generation of carriers through inter-layer photon absorption.	
		(d) The incoming photons of 610 and 613 nm couple with the $ X_{B,D}^- $ state to form the $XX_{B,D}^-$. This many body state acts like an intermediate state that carries a net charge from top FLG film to the bottom. The fat particle $XX_{B,D}^-$ reduces the probability of bidirectional movement and is dragged by the built-in vertical field, contributing to the $I_{ph}$.}\label{fig:mechanism}
\end{figure}

Figure \ref{fig:Iph}a shows the schematic diagram of the 1L-WS$_2$ device (GWG) encapsulated with a thick few-layer graphene (FLG) from top and a relatively thin FLG film from the bottom (see \textbf{Methods} for fabrication details). The optical image of two such devices after completion of fabrication is shown in \textbf{Supporting Information S1}. The different thickness values of the top and bottom FLG layers, and their different dielectric environments induce a difference in doping between the two. This effect, coupled with the ultra-thin nature of the 1L-WS$_2$, results in a strong vertical built-in field in the WS$_2$ layer \cite{kallatt2019interlayer,kim2014high}. Upon photoexcitation, this configuration allows us to obtain a photocurrent ($I_{ph}$) under zero external bias ($V_{ext}$), which has two advantages. First, the average dark current is zero, which in turn suppresses the dark noise. Second, this allows an operation in the self-powered mode without requiring any external power supply to the device. We excite the sample using a tunable source (supercontinuum laser with tunable filter from NKT Photonics), and record both $I_{ph}$ and the photoluminescence (PL) at $4.5$ K (see \textbf{Methods} for characterization setup). Note that the photocurrent away from the GWG junction is significantly suppressed (see \textbf{Supporting Information S1}).
\\
\\
We plot the measured photocurrent ($I_{ph}$ in symbols) as a function of the excitation wavelength ($\lambda$) in Figure \ref{fig:Iph}b. The baseline corresponds to the zero average dark current. The value of $I_{ph}$ is small when $\lambda>620$ nm, suggesting that the photoresponse purely due to graphene absorption is small. The solid trace fitting the data can be decomposed into several individual peaks (shown in shaded areas). To understand the origin of the $I_{ph}$ peaks, we compare them with the peaks obtained in the PL spectra from both the GWG portion as well as an isolated WS$_2$ on hBN (HW) portion (see Figure \ref{fig:Iph}c-d). From the comparison, we identify that the $I_{ph}$ peaks originate from photo-absorption by trion ($X^-$) around $605$ nm and charged biexciton ($XX_D^-$) around $610$ nm. Strikingly, $I_{ph}$ spectrum does not exhibit any peak when the excitation is at the neutral exciton ($X^0$) resonance around $595$ nm \cite{kallatt2019interlayer}. We also observe another strong peak ($XX_B^-$) of $I_{ph}$ around $613$ nm, which is observable both in the PL spectra from the GWG stack as well as the HW stack. We note here that the small spectral shift of the PL peaks between the GWG and the HW stacks results from Coulomb screening which causes a reduction in both the binding energy of the excitonic complexes as well as the quasiparticle bandgap \cite{ugeda2014giant,raja2017coulomb,gupta2017direct}.
\\
\\
To further identify the origin of these peaks, we measure $I_{ph}$ as a function of the incident laser power ($P$) at every wavelength, and fit the obtained data as $I_{ph}\propto P^\alpha$ (Figure \ref{fig:Iph}e). While the $X^-$ peak shows a power-law exponent that is slightly less than unity ($\alpha=0.82\pm 0.05$), the $XX_D^-$ and $XX_B^-$ peaks exhibit $\alpha=1.65\pm 0.08$ and $\alpha=1.8\pm 0.2$, respectively. Under quasi-equilibrium situation, we expect that $\alpha \approx 2$ for the charged biexciton (shown later), which reduces as the system is pushed towards non-equilibrium \cite{chatterjee2021probing}. Thus, the strong non-linearity in the observed $I_{ph}$ with the incident optical power suggests a biexcitonic origin of the two photocurrent peaks. In \textbf{Supporting Information S2}, we obtain similar results from another fabricated device, with $\alpha=1.17\pm 0.16$, $1.41\pm0.14$, and $1.83\pm0.30$ for the $X^-$, $XX_D^-$, and $XX_B^-$ species, respectively. We identify the configuration of the two charged biexcitons as depicted in Figure \ref{fig:Iph}f (supported further later). Note that the observation of multiple biexcitonic peaks have been reported previously as well \cite{zinkiewicz2021excitonic}.
\\
\\
Note that, while such charged biexcitons have been studied in the past through photoluminescence measurements\cite{ye2018efficient,chen2018coulomb,zinkiewicz2021excitonic,barbone2018charge,steinhoff2018biexciton,hao2017neutral,chatterjee2021probing}, no direct evidence about their electronic configuration exists in literature. To explore such configuration of these biexcitonic species, we perform both one- and two-color photoluminescence excitation (PLE) spectroscopy on the isolated part of the WS$_2$ flake. First we scan the excitation wavelength ($\lambda$) and measure the PL intensity at the $XX_D^-$ and $XX_B^-$ peaks at $4.5$ K, as schematically shown in Figure \ref{fig:PLE}a (see \textbf{Methods} for characterization details). The bottom panel of Figure \ref{fig:PLE}b shows that both the peaks exhibit strong emission when the excitation is in resonance with the $X^0$ peak, with the intensity of the $XX_D^-$ being higher than the $XX_B^-$ peak (for reference, the corresponding PL of WS$_2$ is shown in the top panel of Figure \ref{fig:PLE}b). However, luminescence intensity is negligible for both the peaks when the excitation resonates with the $X^-$ peak. This suggests that generating the neutral exciton is crucial to the formation of the charged biexciton, which is otherwise suppressed due to lack of formation of $X^0$ when only the trion (at lower energy) is generated. Further, by keeping the excitation source at the peak intensity wavelength (that is, resonant to $X^0$), we vary its power and record the corresponding PL intensity of $XX_{B,D}^-$ (Figure \ref{fig:PLE}c). The power dependent PL intensity again follows a power-law with $\alpha=1.62 \pm 0.04$ and $1.96 \pm 0.05$ for $XX_D^-$ and $XX_B^-$ peaks, respectively, which are in good agreement with the $I_{ph}$ power-law.
\\
\\
To further verify the nature of the charged biexcitons, we next use a two-color excitation spectroscopy where we keep one excitation wavelength ($\lambda_1$) fixed at $X^0$ resonance and scan the wavelength ($\lambda_2$) of the other excitation, as schematically shown in Figure \ref{fig:PLE}d (see \textbf{Methods}). We observe in Figure \ref{fig:PLE}e that, for the $XX_D^-$ peak, the strongest intensity remains when $\lambda_2$ also resonates with the $X^0$ peak. However, we also observe a non-zero luminescence intensity when $\lambda_2$ is around the $X^-$ peak, unlike the previous single wavelength PLE experiment. On the other hand, for the $XX_B^-$ peak, we notice distinct peaks when $\lambda_2$ is close to both $X^0$ and $X^-$ resonance, both being comparable in intensity (shown in Figure \ref{fig:PLE}e). In particular, the formation of the $XX_B^-$ through the simultaneous absorption of $X^0$ (through $\lambda_1$ in green) and $X^-$ (through $\lambda_2$ in red) is schematically shown in Figure \ref{fig:PLE}d. These observations point to the fact that both $XX_D^-$ and $XX_B^-$ species originate from a combination of $X^0$ and $X^-$, suggesting their charged biexcitonic nature. This is in good agreement with the gate voltage dependence of the PL intensity of the two peaks, measured with a $532$ nm continuous wave laser excitation (See \textbf{Supporting Information S3}). With an increase in the positive or negative gate voltage, the intensity of both the peaks increases due to gate voltage induced doping. This is in good agreement with previous reports as well \cite{ye2018efficient,barbone2018charge}, and supports their charged biexcitonic nature.
\\
\\
Based on the above observations, we propose that $XX_D^-$ is formed through the combination of a bright neutral exciton ($X_B^0$) and a dark trion ($X_D^-$), as reported previously \cite{ye2018efficient,barbone2018charge,chatterjee2021probing} (see left panel of Figure \ref{fig:Iph}f). On the other hand, $XX_B^-$ originates from $X_B^0$ and a bright trion ($X_B^-$), as shown in the right panel of Figure \ref{fig:Iph}f. The long-lived nature of $X_D^-$ results in a higher population density of $XX_D^-$, and hence a stronger PL intensity, compared with $XX_B^-$. As indicated by the dashed lines in Figure \ref{fig:PLE}a, the bright excitons, upon generation, relax to the lower energy dark states by undergoing inter-valley scattering. This explains the reason for $XX_D^-$ showing a strong intensity when $\lambda$ is close to the neutral exciton resonance. This intensity reduces for the $XX_B^-$ case due to quick loss of $X_B^0$ to the lower energy dark states.
\\
\\
Interestingly, while the $XX_B^-$ is of higher energy than $XX_D^-$ , the corresponding transition energy (for both absorption and emission) is less for $XX_B^-$ (Figure \ref{fig:Iph}b-d). This is schematically explained in Figure \ref{fig:PLE}f. The configuration of the two different charged biexcitons in Figure \ref{fig:Iph}f indicates that $E_{X_B^-}-E_{X_D^-}=\Delta_c+\chi$, and $E_{XX_B^-}-E_{XX_D^-}=\Delta_c+\chi^\prime$, where $\Delta_c$ and $\chi$ ($\chi^\prime$) denote the spin-orbit splitting in the conduction bands, and the exchange energy of the trion (charged biexciton) state. Due to an enhanced screening, we expect a reduced wave function overlap in the charged biexciton compared with the trion, and hence $\chi^\prime < \chi$. The transition energy difference between the two species corresponds to $\hbar\omega_{XX_{D}^-} - \hbar\omega_{XX_{B}^-} = \chi - \chi^\prime$, which is $\sim 12$ meV.
\\
\\
After establishing that the two strong photoluminescence peaks around $610$ and $613$ nm originating from $XX_D^-$ and $XX_B^-$, respectively, we now focus on the mechanism that causes the $I_{ph}$, as explained in Figure \ref{fig:mechanism}. Note that, the top and bottom FLG layers can absorb the light resulting in injection of hot electrons and holes from both the FLG layers to WS$_2$. However, due to the large band offset between WS$_2$ and FLG on both sides of WS$_2$, these hot carriers can move towards either direction, preventing any significant net flow of charge along one direction (Figure \ref{fig:mechanism}a). This suppresses the net $I_{ph}$ purely due to FLG absorption, as observed for $\lambda>620$ nm in Figure \ref{fig:Iph}b. We observe that photon absorption by multi-particle states in WS$_2$ clearly enhances $I_{ph}$. As the $I_{ph}$ is obtained under zero external bias, it relies on the vertical built-in field in WS$_2$. However, note that the built-in field is not strong enough to dissociate the strongly bound exciton and other higher order states into individual electrons and holes. The photocurrent is thus suppressed when the excitation is in resonance with the charge neutral exciton (see Figure \ref{fig:Iph}b).

\begin{figure}[!hbt]
	\centering
	\includegraphics[scale=0.5]{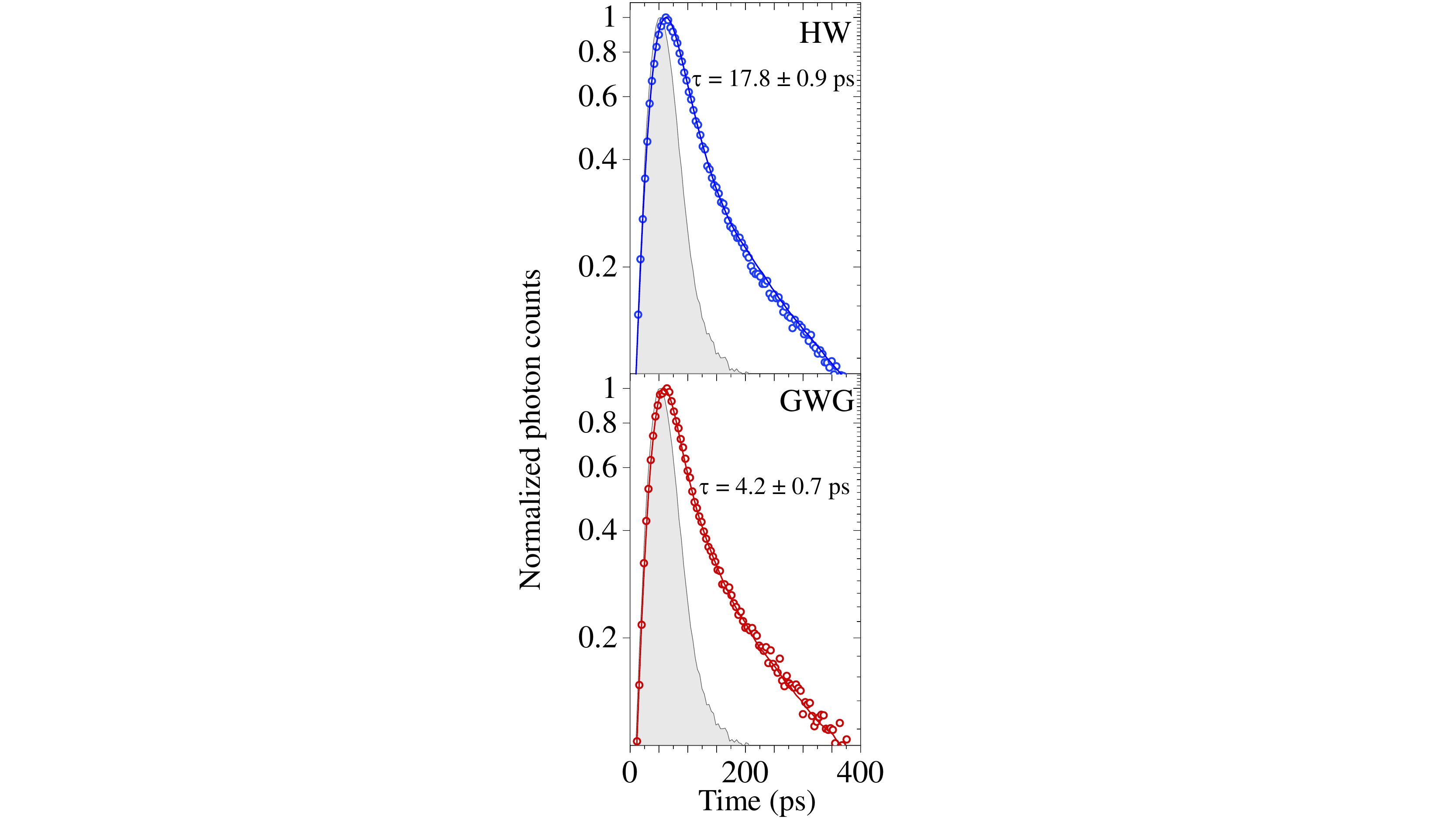}
	\caption{\textbf{Time-resolved photoluminescence measurement.}
		Lifetime of the charged biexciton in (a) HW and (b) GWG stacks measured via time-resolved photoluminescence (TRPL). While the decay time is $17.8 \pm 0.9$ ps on the HW stack, it is $4.2 \pm 0.7$ ps on GWG stack, indicating a faster nonradiative decay of the charged biexciton resulting from inter-layer transfer.}\label{fig:TRPL}
\end{figure}

We first ask the question: what is generation mechanism of the charged biexcitons through resonant absorption of photons that is responsible for the observed $I_{ph}$? There are two possibilities, namely (1) two-photon absorption (TPA) \cite{kamada1998excited,yamada2004stokes,yamamoto2001biexciton}, and (2) absorption of a photon that promotes an already prepared bright or dark trion state to the corresponding charged biexciton state. The TPA process is schematically shown in Figure \ref{fig:mechanism}b. We expect that the two-photon absorption, which occurs through a virtual state (indicated by the dashed line in Figure \ref{fig:mechanism}b), is a relatively weak process. In addition, energy conservation dictates that in the TPA process, the photo-absorption must occur at an energy equal to $0.5(\hbar\omega_{XX_{B,D}^-} + \hbar\omega_{X_{B,D}^-})$. However, our experimental observation suggests that the $I_{ph}$ peaks almost exactly match with the charged biexcitons emission peaks. We thus conclude that the TPA process is not the primary mechanism for the observed photocurrent, and the latter mechanism plays a more important role.
\\
\\
Since charged biexciton transition energy ($\hbar\omega_{XX_{B,D}^-}$) is lower than the trion transition energy ($\hbar\omega_{X_{B,D}^-}$), we rule out the generation of a trion via direct absorption by the WS$_2$ layer at an excitation energy resonant with the $\hbar\omega_{XX_{B,D}^-}$. Instead, hot electrons and holes are generated in the top and bottom FLG layers at these excitation wavelengths due to broadband absorption arising from broad joint density of states in FLG \cite{bassani1967band,johnson1973optical,pedersen2003analytic}. These hot carriers are injected into the WS$_2$ layer through ultrafast inter-layer charge transfer \cite{massicotte2016photo,yuan2018photocarrier,fu2021long}. Note that, while the electron-hole recombination time is short in FLG \cite{george2008ultrafast,rana2009carrier}, the fast out-of-plane transfer timescale allows such hot carrier transfer to WS$_2$ before they recombine in the FLG films \cite{murali2019highly}. These non-equilibrium hot electrons and holes occupy the bands of both valleys in WS$_2$, helping to prepare the initial bright and dark trion state (left panel of Figure \ref{fig:mechanism}c). The excitation pulse width ($\sim 60$ ps) is much larger than the inter-layer charge transfer. Hence, another photon absorption (within the same pulse) with an energy equal to $\hbar\omega_{XX_{B,D}^-}$ promotes the respective trion state to the corresponding charged biexciton state ($XX_{B,D}^-$), as schematically shown in Figure \ref{fig:mechanism}d. We must note that these two processes can happen resonantly as well, which reduces the effective energy barrier for carrier injection from FLG to WS$_2$ due to the Coulomb interaction arising from the many-body effect.
\\
\\
As argued earlier, the hot electrons and holes injected from top and bottom FLG to WS$_2$ do not individually contribute to a large $I_{ph}$ due to a lack of strong asymmetry (Figure \ref{fig:mechanism}a). However, when the injected electron or hole is bound to the fat many-body particle, the probability of movement towards either side is suppressed. Instead, the charged many-body particle itself is dragged by the built-in field, helping in generating a net photocurrent (see Figure \ref{fig:mechanism}d). In addition, when the incoming photon is in resonance with one of the charged many-body states (like trion or charged biexciton), the hot carriers while being injected from FLG, ``feel" the many-body Coulomb interaction, and encounters a smaller injection barrier height, resonantly forming the many-body state.Thus the many-body state acts like an intermediate state that carries the net charge from the top FLG to the bottom FLG film, thus allowing the flow of a net photocurrent conserving the charge.
\\
\\
The right panel of Figure \ref{fig:mechanism}c schematically illustrates another possible way the initial bright or dark trion state can be prepared is by direct inter-layer photon absorption\cite{yuan2018photocarrier}, where an electron in the valence band of WS$_2$ (FLG) is promoted to the conduction band of FLG (WS$_2$), leaving behind a hole. Our experimental observations do not distinguish between these two different mechanisms of initial trion state preparation, and both processes can contribute to the measured $I_{ph}$.
\\
\\
In order to establish the role of FLG in supplying the carriers to form the charge biexcitonic states, we fabricate a lateral hBN capped WS$_2$ device, and measure the lateral photocurrent by exciting the laser spot at the middle of the channel (see \textbf{Supporting Information S4}). The first striking observation is that the total photocurrent is nearly $1000$-fold less in the lateral device, compared with the vertical GWG device. Second, the total photocurrent at the exciton and charged biexciton excitation are similar, unlike the GWG vertical device. Such strong relative suppression of $I_{ph}$ at the charged biexciton excitation energy suggests that the formation of charged biexciton is less efficient without the presence of the FLG.
\\
\\
The two-step mechanism of the generation of the biexcitonic current in the GWG device presented above can be modelled using a simple rate equation:
\beq\label{eq:re}
\frac{dn_{XX^-}}{dt} = \gamma{n_{X^-}}P - \frac{n_{XX^-}}{\tau}
\eeq
Here, $n_{XX^-}$ and $n_{X^-}$ are respectively the charged biexciton and trion population, $P$ is the optical power, $\gamma$ is a proportionality constant, and $\tau$ is the net lifetime of the charged biexciton in the GWG stack. The first term on the right hand side in Equation \ref{eq:re} takes care of the initial trion state formation followed by promotion to a charged biexciton state. From the top panel of Figure \ref{fig:Iph}(e), we note that $n_{X^-} \propto P^{0.82}$. Hence, using $\frac{dn_{XX^-}}{dt}=0$ under steady state, we obtain
\beq
I_{ph} \propto n_{XX^-} \propto P^{1.82}
\eeq
which is in good agreement with the superlinear power law observed at the charged biexciton resonance in the middle and bottom panels of Figure \ref{fig:Iph}(e).
\\
\\
To estimate the timescale of the photocurrent generation mechanism, we compare the net lifetime of the charged biexciton in HW and GWG stacks by measuring time-resolved photoluminescence (TRPL) of the charged biexciton peak (see \textbf{Methods} for measurement details). The results are provided in Figure \ref{fig:TRPL}. In both cases, we are able to fit the charged biexciton population ($n$) decay using two exponentials: $n(t)=A_1e^{-t/\tau}+A_2e^{-t/\tau_d}$, with $A_1/A_2$ being about an order of magnitude. The weaker component arises from the tail of the defect peak, with a slow timescale of $\tau_d$, as observed in the PL spectrum from the GWG stack in Figure \ref{fig:Iph}c. We focus on the stronger (and faster) decay component here, which varies from $\tau=17.8\pm0.9$ ps on the HW stack to $\tau=4.2\pm0.7$ ps on the GWG stack. The net lifetime ($\tau$) of the charged biexciton can be written as
\beq\label{eq:tau}
\frac{1}{\tau}=\Gamma_{r} + \Gamma_{nr}
\eeq
where $\Gamma_r$ and $\Gamma_{nr}$ are respectively the radiative and non-radiative decay rates. Due to graphene induced screening, $\Gamma_r$ is expected to be reduced in the GWG stack compared to the HW stack \cite{gupta2021observation}. Hence, $\frac{1}{\Gamma_r}>17.8$ ps in the GWG stack, which in turn indicates $\frac{1}{\Gamma_{nr}}<5.4$ ps (from Equation \ref{eq:tau}). Note that, $\Gamma_{nr}$ is primarily dominated by the ultrafast inter-layer charge transfer in the GWG stack \cite{lorchat2020filtering,gupta2021observation,fu2021long}, and in turn, is responsible for the photocurrent. The strong reduction of $\tau$ in the GWG stack compared with the HW stack is thus clearly due to such fast inter-layer charge transfer induced enhanced $\Gamma_{nr}$. Such enhanced $\Gamma_{nr}$ in the GWG stack is further supported by the simultaneous observation of the reduced PL counts and a reduced net lifetime of the charged biexciton. This analysis thus suggests that the generated charged biexciton transfers to graphene in a timescale of $<5.4$ ps, indicating the ultrafast nature of the intrinsic photocurrent generation. Note that, the mobility and dimensions of the lateral few-layer graphene access regions, and the contact resistance at the interfaces between the few-layer graphene and metal pad contribute to the parasitic resistance of the device. Thus, while the intrinsic response is in picoseconds, the photocurrent could be eventually limited by the time constant of the entire circuit arising from the parasitic resistance and capacitance.
\\
\\
\section*{Conclusions}
In summary, we demonstrate strong photocurrent generation under zero external bias from resonant absorption by two different five-particle charged biexciton states, followed by their ultrafast inter-layer transfer. The biexcitonic photocurrent is highly nonlinear (near-square-law) with the incident optical power and is significantly stronger than the photocurrent resulting from resonant absorption by the neutral exciton and trion. Further, the nature of the different biexcitonic species are revealed by an interesting single- and two-color photoluminescence excitation scans. Such ultrafast, narrow-band, nonlinear photodetection is promising for several applications such as higher harmonics generation of the modulating signal, receiver design in microwave photonics and visible light communication, square-law circuits, and also in nonlinear optoelectronic applications such as neural circuits. In addition, the techniques used here have intriguing prospects towards alternative spectroscopic techniques for exploring and manipulating many-body states in monolayer semiconductors and beyond.
\\
\\
\section*{Methods}
\textbf{Device fabrication:} We prepare the GWG heterostructure devices using sequential dry-transfer method where the individual layers were exfoliated from bulk crystals onto Polydimethylsiloxane (PDMS) using Nitto tape. The entire stack of FLG-capped monolayer WS$_2$ is made on top of Si substrate covered with 285 nm SiO$_2$. The pre-patterned metal electrodes are prepared using photolithography followed by sputtering of Ni/Au (10/50 nm) and lift-off. The deterministic transfer of layers using micromanipulator ensures that one end of the graphite layers is placed on top the metal pads. After completion of the transfer process, the devices are annealed inside a vacuum chamber (10$^{-6}$ mbar) at 200$^\circ$C for 3 hours for better adhesion of the layers.
\\
\textbf{Scanning photocurrent and PLE measurement:} For the photocurrent measurement, we excite the sample using a supercontinuum laser source cascaded with a multi-channel acuosto-optic tunable filter (NKT Photonics). The laser has a repetition rate of $78$ MHz, and the individual pulses have a width of $\sim 60$ ps. The light is focused on top of the heterojunction using a $\times$50 long-working-distance objective (numerical aperture of 0.5) through an optical window while keeping the device inside a closed-cycle cryostat chamber at $4.5$ K. The power values reported here are measured before the objective using a commercial silicon photodetector (from Edmund Optics), \txb{which has a linear response in the range of the measured power values.} We record the photocurrent under zero external bias using a Keithley 2636B SMU. The photoluminescence excitation (PLE) spectroscopy is performed by exciting the sample with the same tunable source \txb{(linewidth $\sim 0.5$ nm)}. The emitted photons are directed to a spectrometer after passing through a bandpass filter with the central wavelength at $610$ nm and an FWHM of 10 nm. For the two-color excitation, we select two wavelengths from the tunable filter by choosing two different channels, and other components of the setup remain the same as the single-wavelength excitation. A schematic diagram of the measurement setup is shown in \textbf{Supporting Information S5}.
\\
\textbf{TRPL measurement:} We use a 531 nm pulsed laser (PicoQuant) controlled by the PDL 800-D driver (with a pulse width of $\sim 40$ ps and a repetition rate of 10 MHz) for the TRPL measurement. The emission signal is connected to a single photon counting avalanche detector from Micro Photon Devices (SPD-050-CTC). The detector is further connected to a Time-Correlated Single Photon Counting (TCSPC) system (PicoHarp 300 from PicoQuant). We have used a combination of a 600 nm longpass filter and a 610 nm bandpass (FWHM 10 nm) to collect the emitted photon from the charged biexciton states. The FWHM of the instrument response function (IRF) is $52$ ps. We perform the deconvolution using QuCoa software (PicoQuant) that allows us to accurately extract timescale down to $10\%$ of the IRF width \cite{becker2005advanced}.
\section*{SUPPLEMENTARY INFORMATION}
The Supporting Information is available free of charge at XXX. \\
Optical image and low temperature photoluminescence; photocurrent from device-2; gate dependent photoluminescence of 1L-WS$_2$ at $T = 4.5$ K; photoluminescence and photocurrent characteristics from a lateral device; schematic of experimental setup.
\section*{ACKNOWLEDGMENTS}
K.M. acknowledges useful discussion with Varun Raghunathan. This work was supported in part by a Core Research Grant from the Science and Engineering Research Board (SERB) under Department of Science and Technology (DST), a grant from Indian Space Research Organization (ISRO), a grant from MHRD under STARS, and a grant from MHRD, MeitY and DST Nano Mission through NNetRA. K.W. and T.T. acknowledge support from the Elemental Strategy Initiative conducted by the MEXT, Japan (Grant Number JPMXP0112101001) and  JSPS KAKENHI (Grant Numbers 19H05790, 20H00354 and 21H05233).
\section*{Author Contribution}
K.M. designed the experiment. S.D. and S.C. fabricated the devices. K.W. and T.T. provided the hBN flakes. S.D., G.G., and K.M. performed the measurements. S.D., G.G., S.C. and K.M. contributed to the analysis of the data and writing of the paper. S.D. and G.G. contributed equally to this work.
\section*{Competing Interests}
The Authors declare no Competing Financial or Non-Financial Interests.
\section*{Data Availability}
Data available on reasonable request from the corresponding author.

\bibliography{references}
\newpage

\newpage

\newpage

\newpage
\AtEndDocument{

\section*{Supplementary material}

## Supporting Information S1

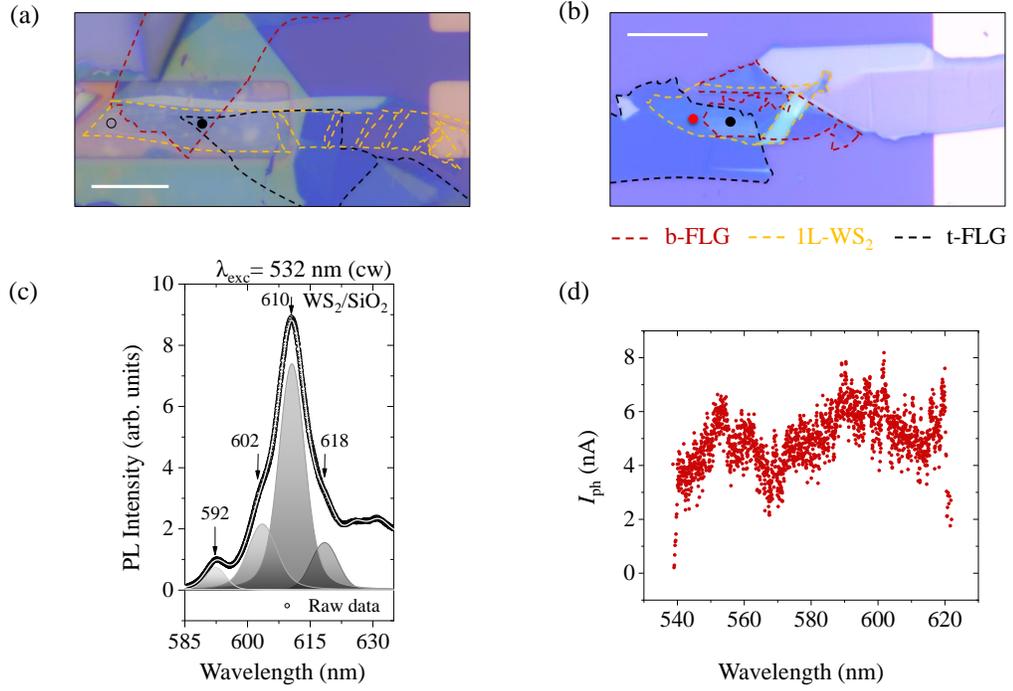

Figure 1: **Optical image and low temperature photoluminescence.** (a-b) Optical image of (a) device 1 and (b) device 2. The experimental results obtained from device 1 are presented in the main text. The experimental results from device 2 are shown in **Supporting Information S2**. The black dots represent the GWG portion in both the devices, and the open circle in (a) represents the HW portion. Boundary of the different materials are delineated with the dashed lines. Scale bar is 5 $\mu$m. (c) PL spectra at $T = 4.5$ K for 1L-WS$_2$ on SiO$_2$ indicating the $X^0$, $X^-$, and $XX^-_{B,D}$ peaks. (d) $I_{ph}$ away from the junction [the location denoted as red dot in (b)] from device 2.

## Supporting Information S2

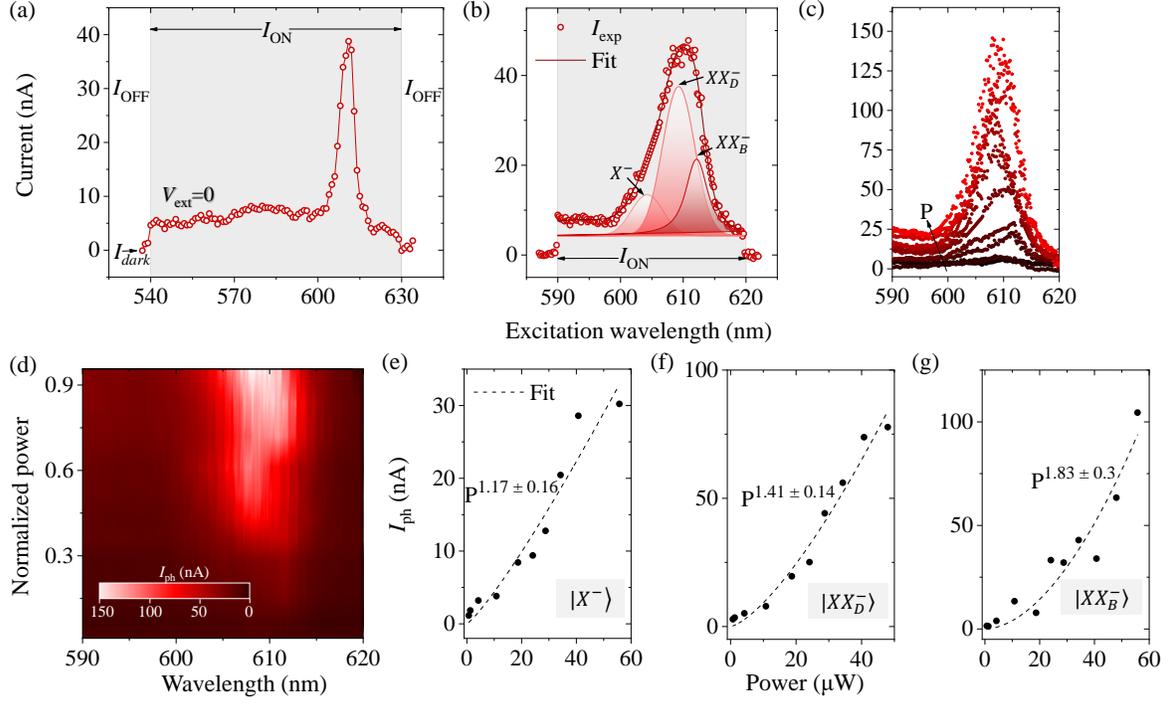

Figure 2: **Photocurrent ($I_{ph}$) from device-2.** (a) The photocurrent spectroscopic response of the device. The shaded region represents the ON current ($I_{ON}$) at zero external bias within the spectral range of $\lambda = 540 - 630$ nm. (b) Photocurrent spectral envelop (shown in open symbol) is decomposed into different energy states as $X^-$, $XX_D^-$ and $XX_B^-$. (c) $I_{ph}$ as function of increasing power ($P$) within the spectral range shown in b. (d) 2-D color plot of the $I_{ph}$ within the spectral range. (e-g) Plot of $I_{ph}$ power law for the $X^-$, $XX_D^-$ and $XX_B^-$ states in (e), (f) and (g) respectively. The dashed lines correspond to the fit of $I_{ph} \propto P^\alpha$. The $\alpha$ values are in good agreement with the device-1.

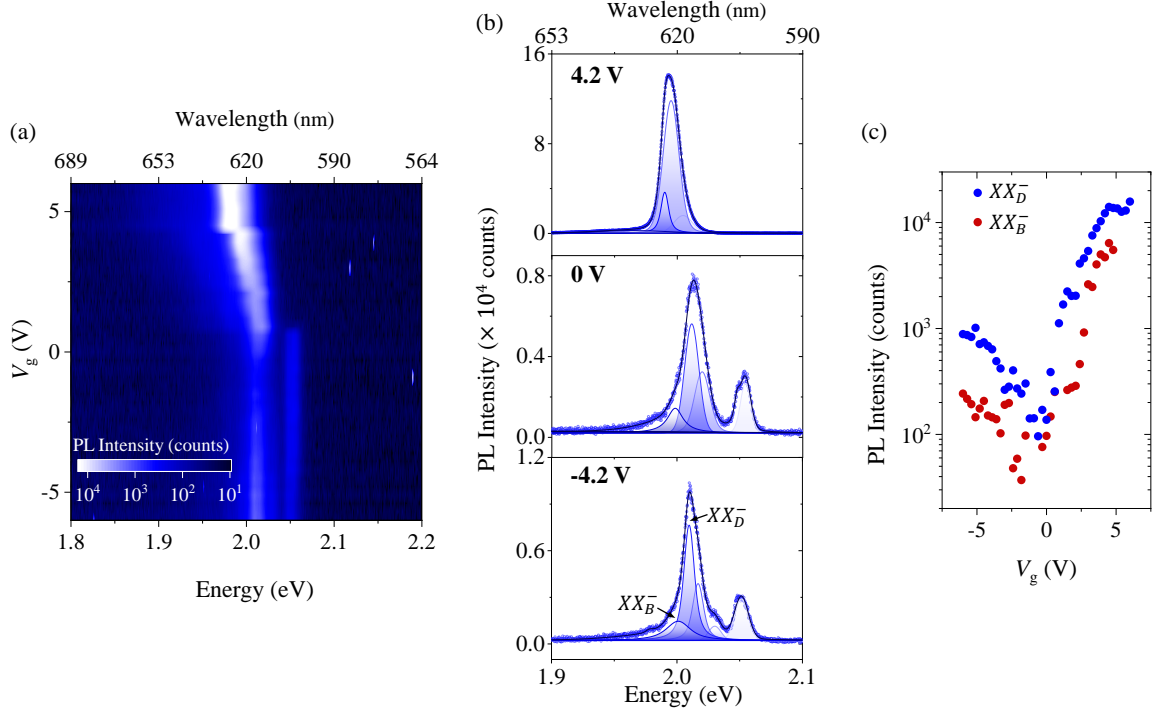

Figure 3: **Gate dependent photoluminescence of 1L-WS$_2$ at** $T = 4.5$ **K.** (a) color plot of the gate dependent PL intensity of 1L-WS$_2$ at $T = 4.5$ K. (b) Line cut of PL spectra at three different voltage points ($V_g = 4.2, 0, -4.2$ V). The spectra are further fitted with voigt function to extract the individual energy states ($X^0$, $X^-$, $XX_D^-$ and $XX_B^-$). While the $X^0$ peak is present at neutral region, it vanishes at the positive $V_g$ or in the $n$-doped region. The $XX_{B,D}^-$ energy states are highly tunable across the doping regions. (c) PL intensity of the $XX_{B,D}^-$ state extracted from the fitting shown in b. The intensity tunability with gate voltage confirms the charged nature of both the species.

# Supporting Information S4

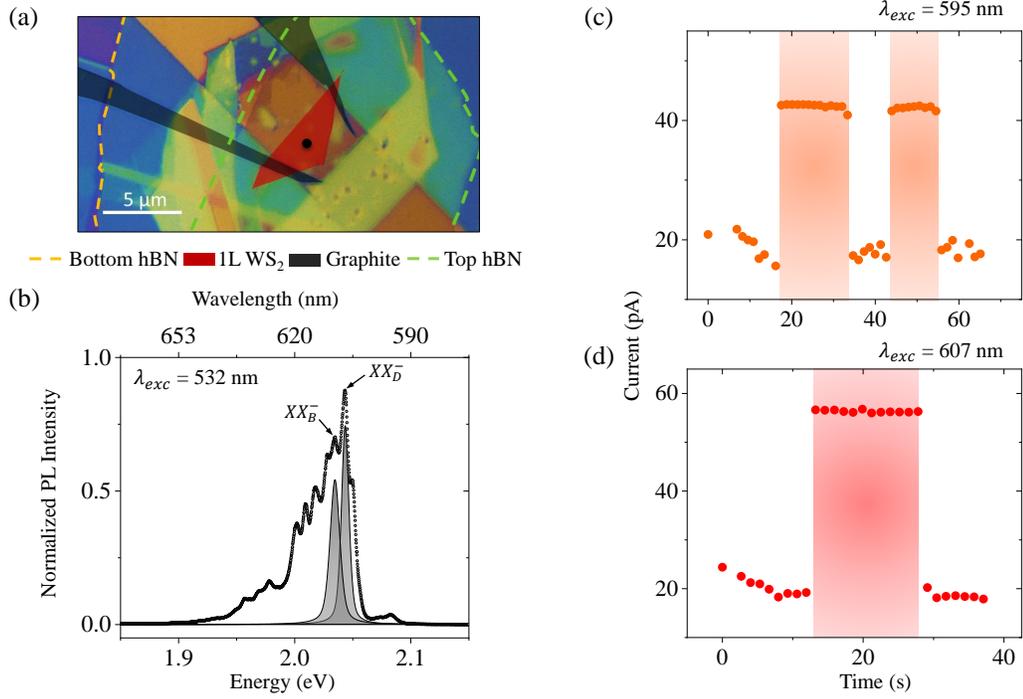

Figure 4: **Photoluminescence and photocurrent characteristics from a lateral device.** (a) Optical image for the lateral device of 1L-WS$_2$ capped with hBN. (b) PL spectrum of hBN capped 1L-WS$_2$ at $T = 4.6$ K with 532 nm cw laser excitation. The fitting of the two charged biexcitons ($XX^-_{B,D}$) is shown for reference. (c-d) Current from the lateral device with excitation at (c) $X^0$ resonance ($\lambda_{exc} = 595$ nm) and (d) $XX^-_D$ resonance ($\lambda_{exc} = 607$ nm) with $V_{ds} = 1$ V, and $T = 4.6$ K. The shaded and unshaded regions indicate light on and off, respectively. The corresponding $I_{ph}$ is about 1000-fold lower compared to the vertical device described in the main text. In addition, there is no significant variation in $I_{ph}$ observed upon resonant excitation at neutral exciton or charged biexciton.

Figure 5: **Schematic of experimental setup.** Schematic diagram of the experimental setup used in photocurrent spectroscopy, PLE scan, and TRPL measurements.

}	
\end{document}